\documentclass{optica-article}

\journal{opticajournal} % for journals or Optica Open

\articletype{Research Article}

\usepackage{pdfpages}
\usepackage{lineno}
\begin{document}

\title{Multipolar optical binding in focus}

\author{Ashutosh Shukla\authormark{$\dagger$}, Sneha Boby, and G V Pavan Kumar\authormark{*}}

\address{Department of Physics, Indian Institute of Science Education and Research Pune, Pune, Maharashtra, 411008, India}

\email{\authormark{$\dagger$}ashutosh.shukla@iiserpune.ac.in}
\email{\authormark{*}pavan@iiserpune.ac.in} %% email address is required; see note below about the corresponding author designation

% use {abstract*} to suppress the copyright line. Copyright information will be added in production

\begin{abstract*} 
The optical binding of gold nanoparticles has conventionally been explored within the Rayleigh limit using dipole approximations. But the field is increasingly focusing on the Mie regime for particles in the 100-500 nm range, where the dipole approximation is insufficient, and a complex landscape of multipolar resonances must be considered. This can be leveraged to engineer more complex forms of optical matter. To this end, we computationally study the optical binding force landscapes experienced by a pair of AuNPs using generalised multiparticle Mie theory. We calculate the total optical binding forces and mechanical trap stiffness values ($dF_i/di$) at the specific resonance wavelengths where the electric dipole, quadrupole, or octupole modes reach their respective scattering peaks and dominate the mechanical response. We demonstrate that the plasmonic mode symmetry greatly influences the spatial distribution of zero-force nodes and the rigidity of the optically bound dimer. By aligning these multipolar phenomena with standard experimental configurations, this work provides a mechanical framework for programmable metafluids and reconfigurable micromachines, bridging the gap between fundamental electrodynamics and reconfigurable nanomanipulation. 
\end{abstract*}

% 

%%%%%%%%%%%%%%%%%%%%%%%%%%  body  %%%%%%%%%%%%%%%%%%%%%%%%%%
\section{Introduction}
The field of optical manipulation has evolved significantly since the pioneering demonstration of optical tweezers\cite{ashkinObservationSinglebeamGradient1986,volpeRoadmapOpticalTweezers2023a, ghoshAllOpticalDynamic2019}, expanding from the confinement of single particles to the orchestration of complex, many-body systems\cite{royControlledTransportationMesoscopic2013,sikdarTunableBroadbandOptical2014,sikdarOptimizedGoldNanoshell2013,shuklaSynchronizedMotionGold2025a,chandOptothermalRevolutionColloids2025a,patraPlasmofluidicSingleMoleculeSurfaceEnhanced2014a,huangUnidirectionalOpticalSwarming2023a,kudoSingleLargeAssembly2018a}. While early work focused primarily on dielectric beads, the integration of plasmonic particles, specifically gold nanoparticles (AuNPs), has revolutionized the field. Due to their high polarizability and localized surface plasmon resonance \cite{faradayBakerianLectureExperimental1857,maierPlasmonicsFundamentalsApplications2007,crozierPlasmonicNanotweezersWhats2024a,juanPlasmonNanoopticalTweezers2011b,willetsLocalizedSurfacePlasmon2007,amendolaSurfacePlasmonResonance2017}, these particles allow for the assembly of dynamic structures ranging from linear chains and 2D crystalline lattices to fluid-like swarms in solution\cite{patraPlasmofluidicSingleMoleculeSurfaceEnhanced2014a,karasekAnalysisOpticalBinding2006,yanGuidingSpatialArrangements2013b,taoRotationalDynamicsIndirect2023b,huangPrimevalOpticalEvolving2022a}.  Often these collective dynamics are enabled by optical binding\cite{burnsOpticalBinding1989c,mohantyOpticalBindingDielectric2004c,parkerOpticalMatterMachines2020a,burnsOpticalMatterCrystallization1990,yanPotentialEnergySurfaces2014a,grzegorczykStableOpticalTrapping2006c}. Distinct from the gradient force, which confines a particle to the laser focus, optical binding arises from the scattering of light between particles\cite{dholakiaColloquiumGrippedLight2010b,forbesOpticalBindingNanoparticles2020e}. It is this inter-particle radiative coupling that generates a modulated force landscape, establishing the equilibrium distances, symmetries, and mechanical rigidity necessary for the formation of optical matter. Applications of plasmonic optical matter include the development of reconfigurable optical micromachines, such as light-driven rotors, micro-gears, and soft robotic components.

To enable these applications, we need to understand the properties of optical matter (OM) outside the dipole approximation. A significant portion of existing optical binding literature treats nanoparticles within the Rayleigh approximation, effectively modeling them as simple point dipoles. This assumption holds for small particles but breaks down for AuNPs in the mesoscopic size range ($x=\frac{2\pi r}{\lambda/n_m}\approx 1$, here $250$ to $500$ nm), where the particle diameter becomes comparable to the excitation wavelength. In this Mie regime, phase retardation effects across the particle volume are non-negligible, leading to the excitation of higher-order multipole modes, specifically electric quadrupole and electric octupole resonances, in addition to the fundamental dipole mode. Notably, these higher-order modes exhibit unique scattering symmetries that fundamentally alter the inter-particle interaction force landscape\cite{huangPlasmonicDipoleQuadrupole2024a,huangSurfacePlasmonResonance2020}. Since most experimental studies rely on a single working wavelength (often $1064$ nm) that is far detuned from these resonances, the distinct mechanical roles of specific plasmon modes remain largely unresolved. Studying the granular role of these plasmon resonance modes allows for improved control over the light-matter interaction. By tuning the incident laser wavelength to excite specific multipolar resonances or their combinations, one can modulate the sign of the interaction (attraction vs. repulsion) and the effective range of the binding force. This capability moves the field beyond static trapping, enabling the programmable assembly, disassembly, and reconfiguration of plasmonic structures in water.

In addition to wavelength selection, beam focusing can substantially influence the optical binding landscape. Tight focusing introduces strong field gradients that not only trap particles but can also drive higher-order multipole modes via displacement resonance, which are weakly excited under plane wave illumination\cite{tangMultipoleEngineeringDisplacement2023,smirnovaMultipolarNonlinearNanophotonics2016,zurita-sanchezMultipolarInterbandAbsorption2002a,novotnyPrinciplesNanoOptics2012,dasBeamEngineeringSelective2015,hanGeneralizedLorenzMie2003}. As a result, the observed force landscape can become a hybrid of the gradient forces and multipolar scattering dependent optical binding. 

In this work, we present a systematic computational study of optical binding of gold nanoparticles in the Mie regime, evaluating the total electrodynamic force fields and mechanical trap stiffness metrics. Since the binding is dependent on the scattering properties of the nanoparticles' plasmon modes, we perform our study at different wavelengths where the dominant scattering contribution is from the electric dipolar, quadrupolar, or octupolar plasmon resonance. We demonstrate that the dominant plasmon mode governs the two-dimensional vector force landscapes and trap stiffness configurations ($dF_i/di$) of optically bound particles, thereby dictating their stable binding distance and orientation. Furthermore, we examine transition wavelengths at which two consecutive modes (such as the dipole and quadrupole) possess equal scattering cross-sections and contribute equivalently to the total response. At these wavelengths, phase interference between the co-excited modes significantly reconfigures the stability and landscape of the mechanical traps. Finally, to understand the role of beam focusing in optical binding, we systematically analyze optical binding forces across particle sizes using tight ($1~\mu$m) and weak ($3~\mu$m) focusing beam waists. We compare these forces with those from plane-wave illumination and find that tight focusing amplifies the gradient force, leading to a predominantly attractive interaction for small particles.

\section{Methods}
\subsection{Simulation Geometry}
The optical matter assembly consists of a pair of identical spherical gold nanoparticles of radius $a$ (with diameters $2a$ varying between 100 nm and 500 nm). Both particles are securely confined to the transverse focal plane of the laser beam, $z_1 = z_2 = 0$. Due to the underlying electromagnetic symmetries of the system under $x$-polarized illumination, the mechanical landscape possesses mirror symmetries across both the $x=0$ and $y=0$ Cartesian axes. Consequently, the 2D spatial scans are systematically computed by fixing Particle 2 at the coordinate origin $\mathbf{r}_2 = (0,0,0)$ while scanning the center position of Particle 1 across the positive first quadrant of the $xy$-plane ($6~\mu m \ge \Delta x \ge 2a, 6~\mu m \ge \Delta y \ge 2a$). The lower limit is set slightly more than 2 times the particle radius to prevent unphysical volume intersections and strong near-field optical forces, which make it harder to visualize the far-field optical binding forces. A detailed visualization of the three-dimensional coordinate system, the laser propagation and polarization axes, and the precise grid boundaries enclosing the hard-sphere exclusion zone is provided in Section S1 of the Supporting Information.

\subsection{Theoretical Framework (GMMT) and Vectorial Beam Formulation}
Electromagnetic scattering calculations were performed using the Generalized Multiparticle Mie Theory (GMMT)\cite{xuElectromagneticScatteringAggregate1995}, implemented in the open-source Python library \textit{MiePy}\cite{johnaparkerGitHubJohnaparkerMiepy, parkerOpticalMatterMachines2020a}. Unlike point-dipole approximations, GMMT provides a rigorous, full-wave solution to Maxwell's equations for aggregates of spherical particles by expanding the incident, internal, and scattered fields in a complete basis of Vector Spherical Harmonics (VSH). 

To ensure mathematical and physical rigor in modeling tightly focused beams, the incident fields are constructed using the Angular Spectrum Representation (ASR) \cite{novotnyPrinciplesNanoOptics2012}. Rather than relying on scalar approximations, \textit{MiePy} decomposes the focused laser field by executing a numerical quadrature superposition integral of plane-wave components sweeping across the forward hemisphere up to a sharp cutoff angle ($\theta_{\text{max}} = \pi/2$). This full vectorial ASR treatment captures the non-paraxial phase curvatures, amplitude gradients, and cross-polarized components inherent to localized focus regions, guaranteeing that gradient-dependent displacement resonances and higher-order multipole excitations are rigorously accounted for.

%{\color{red} 
While the angular spectrum representation ensures that the near-field is reconstructed via an exact, Maxwell-consistent plane-wave superposition integral without paraxial propagation errors, we note a fundamental limitation regarding the initial source envelope. Because the standard Gaussian beam ansatz fundamentally originates from the paraxial wave equation, scaling the synthetic beam waist down to $w_0 = 1\ \mu\text{m}$ (where $w_0 \sim \lambda$) pushes the limits of physical validity. In physically realizable high-NA optical trapping systems, focal fields exhibit complex polarization mixing and strong longitudinal field components best described by Richards-Wolf vectorial focusing theory. Therefore, the $w_0 = 1\ \mu\text{m}$ condition simulated here should be interpreted with caution; it serves as a theoretical limit to qualitatively illustrate the onset of gradient-enhanced multipolar effects rather than a quantitatively exact reproduction of a tightly focused experimental beam.%}

\subsection{Electrodynamic Force Evaluation and Trap Stiffness Quantification}
Time-averaged optical forces exerted on the nanoparticles were calculated by integrating the Maxwell Stress Tensor (MST)\cite{bartonTheoreticalDeterminationNet1989} over a closed spherical surface $\Omega$ enclosing each particle:
\[
\langle \mathbf{F} \rangle = \oint_{\Omega} \langle \overline{\overline{\mathbf{T}}}_{\text{MST}} \rangle \cdot d\mathbf{\Omega}
\]
In the computational implementation, this surface integration is evaluated analytically using the coupled far-field expansion coefficients of the interacting scattered and incident fields, providing exceptional numerical stability even when particles approach near-field contact. 

Because electrodynamic scattering systems operating in the Mie regime are inherently subject to radiation pressure, phase retardation, and wave interference, the resulting vector force fields exhibit a non-zero curl. Consequently, to evaluate trapping mechanical properties directly, we map the local force gradients to determine the structural trap stiffness ($k_c$). The principal mechanical restoring spring constants along the Cartesian polarization axis ($x$) and the perpendicular axis ($y$) are extracted via directional numerical differentiation of the spatial force profiles:
\[
k_x = -\frac{\partial F_x}{\partial x}, \quad k_y = -\frac{\partial F_y}{\partial y}
\]
Stable mechanical trapping configurations are strictly identified at spatial coordinates where the net force vector satisfies $\mathbf{F} = 0$ concurrently with negative tracking slopes ($\partial F_i / \partial i < 0$). The multipole expansion truncation order $L_{max}$ was conservatively set to $L_{\text{max}} = 15$ for all calculations, ensuring full numerical convergence. We tested convergence as a function of truncation order, as discussed in Supporting Information S2.

\subsection{Simulation Parameters and Data Analysis}
Simulations were conducted for pairs of gold nanoparticles (AuNPs) with diameters ranging from $100$~nm to $500$~nm. The complex wavelength-dependent dielectric permittivity of gold was incorporated from the experimental data of Johnson and Christy\cite{johnsonOpticalConstantsNoble1972}. The particle pairs were assumed to be completely immersed in an unconfined ambient water medium ($n_m = 1.33$). We systematically investigated the optomechanical binding landscapes under three distinct illumination profiles: an unpolarized plane wave, and $x$-polarized Gaussian beams with beam waists of $w_0 = 3~\mu\text{m}$ (weak focusing) and $w_0 = 1~\mu\text{m}$ (strong focusing). 

The 2D spatial distribution of the optical force components $\mathbf{F}(\mathbf{r})$ was systematically mapped by scanning the inter-particle separation vector across a 2D grid spanning the $xy$-plane with a high-resolution spatial step size of $10$~nm. All post-processing routines, 2D vector quiver visualizations, force field spatial derivations for stiffness mapping, and 1D line force profiles were processed and generated using MATLAB.

\section{Results and Discussion}
There is a compelling need to extend the understanding of optical binding into regimes where higher-order particle multipoles are excited. Accessing these modes requires an increase in particle size, which introduces specific constraints regarding material properties and heating. For gold nanoparticles ranging from 10 nm to 1000 nm, the absorption cross-section ($C_{\text{abs}}$) is substantial in the near-UV and visible regions, peaking at 600 nm. To avoid the detrimental effects of absorption and consequent particle heating, we use laser wavelengths above this high-absorption window, operating where scattering cross-sections ($C_{\text{sca}}$) dominate the total optical response.

To rigorously address the mechanical effects of higher-order modes, we evaluate the total inter-particle force landscapes at specific spectral peaks where a given multipole mode reaches its maximum relative weight. In this Mie regime, pure modal isolation is physically impossible; multiple multipolar orders are excited simultaneously and couple with one another. However, by tracking the scattering cross-section spectrum via GMMT decomposition (as shown in Fig. 1), we identify distinct wavelengths at which a single mode comprehensively dominates the mechanical response. For example, for a 500~nm AuNP pair, the total force landscape evaluated at $\lambda = 1858~\text{nm}$ is designated as the Electric Dipole (ED) regime; at $\lambda = 973~\text{nm}$, the response is governed by the Electric Quadrupole (EQ) peak; and at $\lambda = 703~\text{nm}$, it is driven by the Electric Octupole (EO) peak.

Furthermore, because our methodology calculates the net time-averaged forces holistically through the complete Maxwell Stress Tensor, all concurrent cross-modal interactions and secondary contributions are inherently integrated into the final landscapes. This also includes the minor roles played by magnetic multipoles, such as the magnetic dipole (MD) mode. Given their minimal scaling relative to the massive cross-sections of the dominant electric modes, magnetic contributions exert a negligible influence on the overall mechanical trap geometry and stiffness parameters. Consequently, our discussion focuses on the structural and mechanical transitions dictated by the dominant electric multipole states.

\subsection{Multipolar scattering and optical binding}
\begin{figure}[h!]
\centering\includegraphics[width=\textwidth]{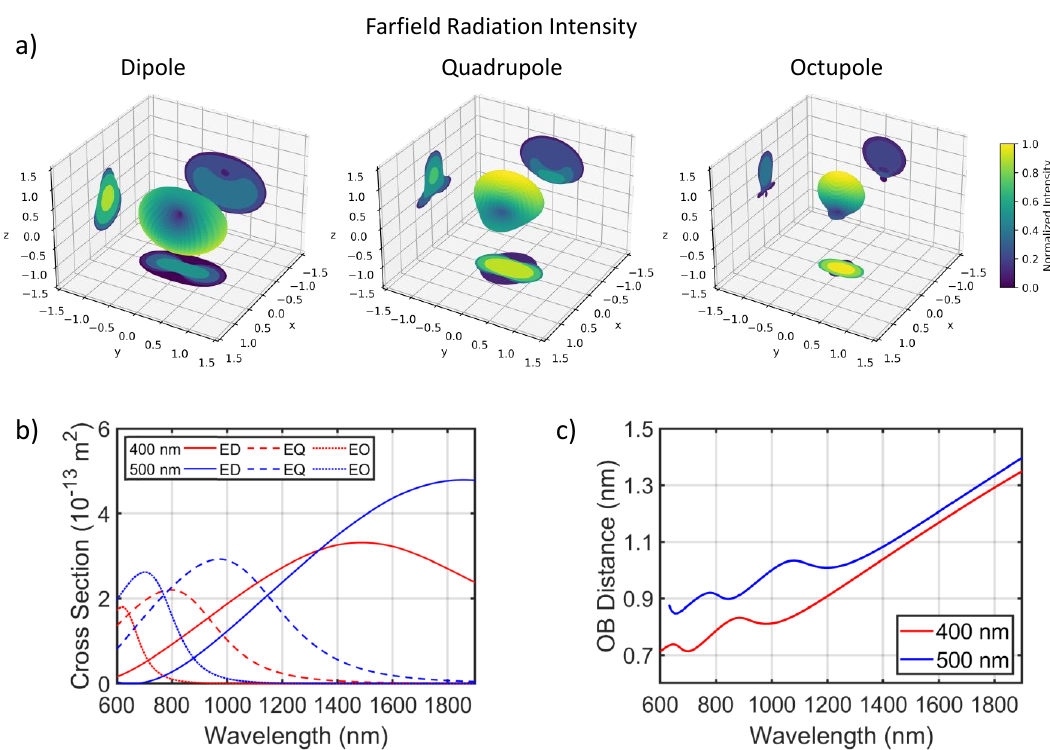}
\caption{Scattering at various resonances and optical binding distance. (a) Total far-field angular scattering intensity for a 400 nm AuNP at different plasmon resonances with projections onto Cartesian planes. (b) Scattering cross-sections of 400 nm and 500 nm AuNPs decomposed into the electric dipole (ED), quadrupole (EQ), and octupole (EO) modes. (c) The stable optical binding (OB) distance for 400 and 500 nm AuNPs is not a linear function of laser wavelength.}
\label{SCS_OBD}
\end{figure}

Our study focuses on a parameter space with particle sizes ranging from 100 nm to 500 nm and laser wavelengths from 600 nm to 1900 nm. This approach allows us to map the total electrodynamic interaction induced force landscapes and mechanical stability of the optically bound dimer when higher-order modes dominate the particle response. Since these quadrupole and octupole modes possess distinct radiation scattering patterns in the three dimensions around the particle, their excitation fundamentally reshapes the spatial distribution of the equilibrium positions for the dimer. We show separately in Figure \ref{SCS_OBD}a the distinct farfield angular scattering patterns of a 400 nm AuNP when its Dipole, Quadrupole, and Octupole modes are excited.

To gain a continuous spectral perspective on the optical binding behavior, we analyzed the scattering properties and the corresponding equilibrium separation distances across the bandwidth of interest ($600 - 1900 \text{ nm}$). Figure \ref{SCS_OBD}b presents the scattering cross-sections for 400 nm (red) and 500 nm (blue) AuNPs, decomposed into their constituent electric dipole (ED), electric quadrupole (EQ), and electric octupole (EO) modes using generalized Mie theory.

The spectra clearly show a redshift of the resonance peaks with increasing particle size. For the 500 nm particle, the octupole, quadrupole, and dipole modes peak at approximately $703 \text{ nm}$, $973 \text{ nm}$, and $1858 \text{ nm}$, respectively. We can quantify the contribution of these modes at their respective peak wavelengths to ascertain their dominant contributions in the scattering field. For the 500~nm particle, at $\lambda = 1858.4$ nm, the ED mode carries a massive fraction of $90.02\%$ ($\sigma_{\text{ED}} = 4.79 \times 10^{-13}\text{ m}^2$) over a minor $8.82\%$ magnetic dipole (MD) component; at $\lambda = 973.3$ nm, the EQ mode governs $50.10\%$ ($\sigma_{\text{EQ}} = 2.92 \times 10^{-13}\text{ m}^2$) of the cross-section; while at $\lambda = 703.4$ nm, the EO mode claims $39.79\%$ ($\sigma_{\text{EO}} = 2.62 \times 10^{-13}\text{ m}^2$) of the total scattering cross section. This while seeming small is still the dominant contribution in the scattering followed by the EQ mode with $24.36\%$ and rest of it divided into the magnetic modes primarily. Similarly, for the 400 nm particle, the octupole, quadrupole, and dipole modes peak at approximately $619 \text{ nm}$, $796 \text{ nm}$, and $1487 \text{ nm}$, with $91.54\%$, $53.15\%$ and $38.41\%$ contributions to the total scattering cross section at those wavelengths respectively. The plot also shows the transition wavelengths discussed previously. Specifically, the wavelengths at which the ED and EQ curves intersect are 1147 nm and 932 nm for 500 nm and 400 nm particles, respectively. Also, the intersection points of the EQ and EO curves lie at 770 nm and 644 nm for 500 nm and 400 nm particles, respectively. These intersection points represent regimes of strong multipolar phase interference, in contrast to the resonance peaks where primarily one mode dominates the scattering profile.

Figure \ref{SCS_OBD}c plots the stable optical binding (OB) distance as a function of wavelength for the same particle sizes. We calculate the force on particles by placing them at increasing interparticle separations perpendicular to the laser polarization. The binding distance is defined here as the separation at which the optical force transitions from repulsive to attractive. While there is a general linear trend in which the binding distance scales with wavelength (roughly $\sim \lambda$), significant deviations occur in the short-wavelength region, where contributions from quadrupolar and higher modes become dominant.

This correlation can also be seen by comparing the two panels: the monotonic increase in binding distance is modulated by the excitation of higher-order modes. The wiggles or oscillations in the binding distance curve in Figure \ref{SCS_OBD}c coincide spectrally with the regions in Figure \ref{SCS_OBD}b where the quadrupole and octupole modes are significant. This confirms that the interplay between multipolar modes does not merely modulate the restoring force magntiude, but actively shifts the spatial position of the stable equilibrium, creating deviations from the standard $\lambda$-scaling typically observed in the dipolar regime.

\subsection{Optical binding force landscapes at different plasmon resonances}
\begin{figure}[h!]
\centering\includegraphics[width=\textwidth]{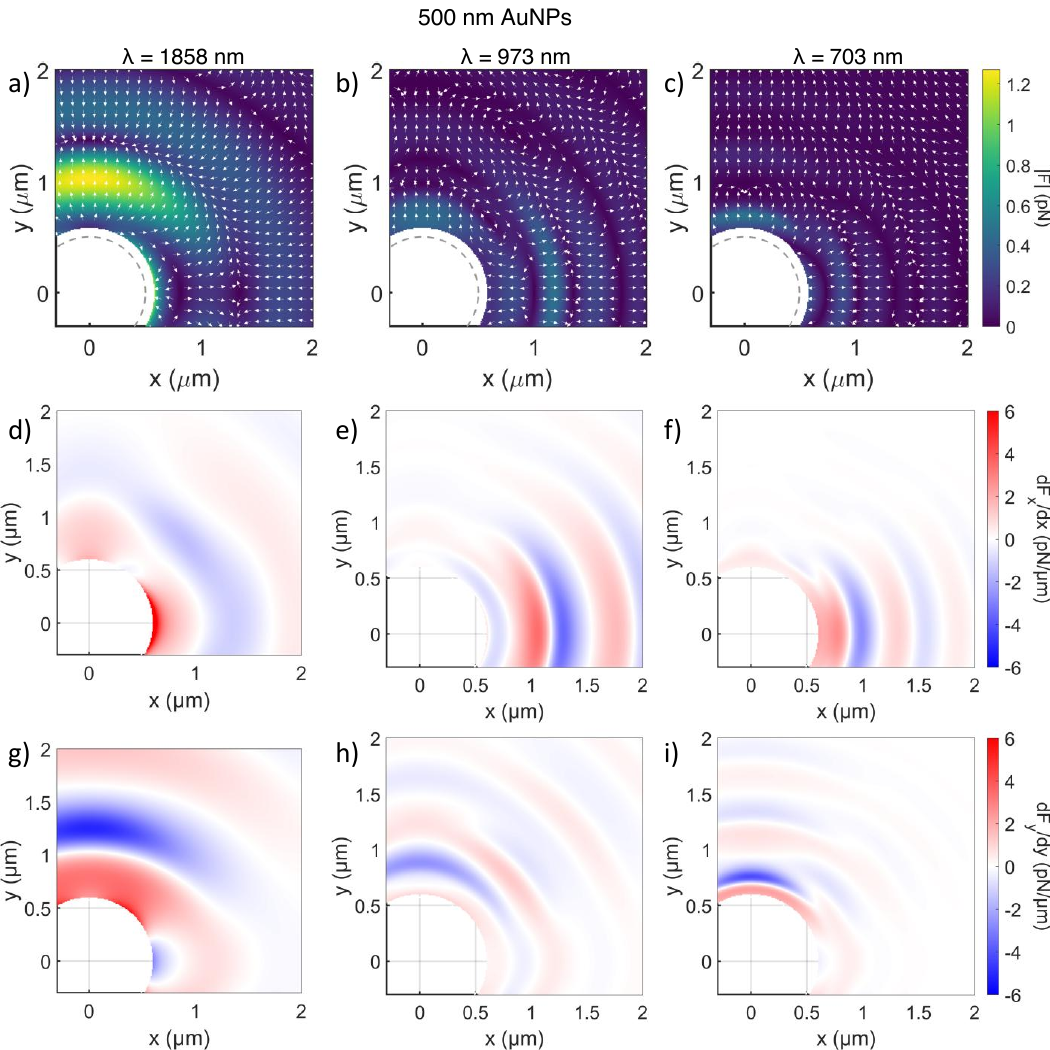}
\caption{Two-dimensional optical force landscape at various plasmon resonance modes for 500 nm AuNP pairs. (a–c) Force vector quiver plots superimposed over the net force magnitude $|F|$ for a pair of 500 nm gold nanoparticles illuminated by an $x$-polarized Gaussian beam ($w_0 = 3~\mu\text{m}$) at wavelengths aligning with the peaks of the (a) Electric Dipole (ED) at $\lambda = 1858~\text{nm}$, (b) Electric Quadrupole (EQ) at $\lambda = 973~\text{nm}$, and (c) Electric Octupole (EO) at $\lambda = 703~\text{nm}$. Due to mirror symmetries across the Cartesian axes, independent features are calculated in the first quadrant and mirrored. (d–f) Corresponding maps of trap stiffness along the $x$-direction ($dF_x/dx$). (g–i) Corresponding maps of trap stiffness along the $y$-direction ($dF_y/dy$). Regions where physical particle overlapping occurs ($r < 2a$) are masked in white.}
\label{5EDQO}
\end{figure}

The optical binding force between two particles is anisotropic, and looking at the two-dimensional force landscape is necessary to understand the full binding behaviour. To achieve this, we scan the particle pair positioned at \( \{(x,y), (0,0)\},\textrm{ where, }x,y\in [0\ \mu m, 6\ \mu m]  \) and calculate the force on the first particle. The step size in both directions is 10 nm. Then, we map the absolute force between the particles at all the positions and extract the local trap stiffness by calculating the diagonal components of the stiffness matrix ($k_x = -\partial F_x / \partial x$ and $k_y = -\partial F_y / \partial y$). Due to the symmetries of the calculation, we constructed the full two dimensional vector force field as shown partly in Figure \ref{5EDQO} by patching the appropriately mirrored parts. The region where particles overlap is manually removed and shown in white. The two-dimensional vector force landscape, generated as described for a 500 nm particle pair illuminated with a 1858 nm laser and a 3 \(\mu\)m beam waist, with polarization along the x axis, is shown in Figure \ref{5EDQO}a. The particular laser wavelength corresponds to the peak of the scattering cross section due to the dipolar plasmon resonance of a 500 nm AuNP. The stable equilibrium positions can be identified in the force field at coordinates where the net force crosses zero with a negative slope ($\mathbf{F}=0$ and $\partial F_i / \partial i < 0$). The particle at $(x,y)$ follows the local force profile to reach an equilibrium position. In the dipolar regime, this position is generally perpendicular to the trapping laser polarization. In our case, the stable equilibrium position is at a center-to-center separation of 1.37 $\mu$m, which is roughly equal to $\lambda/n_m=1.858/1.33=1.39~\mu$m. We show the trap stiffness of the binding along the x and y directions in Figure \ref{5EDQO}d) and g). While an equilibrium position also exists along the x=0 line, it is an unstable equilibrium position along the y-axis as can be seen by comparing Figure \ref{5EDQO}d) and g).

We proceed to calculate the vector force fields and directional gradients at the laser wavelength corresponding to the peak of the quadrupole plasmon resonance of the 500 nm AuNP, which is at 973~nm. It is shown in Figure \ref{5EDQO}b. It is notable that the mechanical landscape differs significantly in shape and confinement properties as shown in Figure \ref{5EDQO}e) and h). Firstly, the multi-lobed scattering profile triggers a fundamental structural reconfiguration of the force field. A new local extremum has emerged oriented in the x-direction, which is also the direction of laser polarization. The equilibrium separation in this case is around $x = 1.350~\mu\text{m}$ ($\partial F_x/\partial x = -2.731\text{ pN/}\mu\text{m}$). This implies that the optical binding morphology of arrays of two or more particles created at the quadrupolar resonance would be significantly different than that at the dipole resonance. Further, we show in Figure \ref{5EDQO}c, the force distribution and directional derivative maps for the same 500~nm AuNPs at their octupole resonance, which is at 703.4~nm. Again, the mechanical topography looks very different from the dipole and quadrupole landscapes. The stable equilibrium position is now aligned perpendicular to the laser polarization, with an interparticle separation at $y = 0.900~\mu\text{m}$.

We show similar vector force fields and directional stiffness arrays for a pair of 400 nm particles at laser wavelengths corresponding to dipolar (1487 nm), quadrupolar (796 nm), and octupolar (619 nm) resonance peaks in Supplementary Information. Notably, the spatial configurations are qualitatively similar to the 500~nm dimer.

In all these cases, the force landscape is periodic, with a period roughly equal to the laser wavelength. The binding landscapes are significantly influenced by the laser wavelength used to optically bind the particles. This is because the angular scattering profile depends on the plasmonic resonance excited in the particle, as shown in Figure \ref{SCS_OBD}a, which is wavelength-dependent. 

\begin{figure}[h!]
\centering\includegraphics[width=\textwidth]{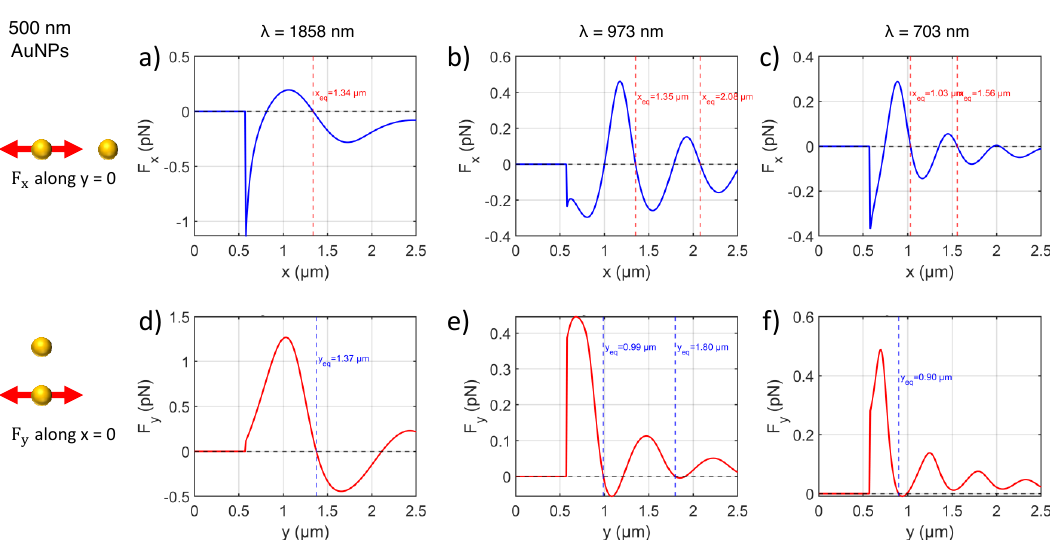}
\caption{One-dimensional line force cuts and mechanical trap configurations for electric multipole peaks. Line profiles extracted for 500 nm AuNP pairs under weak focusing ($w_0 = 3~\mu\text{m}$). (a-c) $F_x$ forces evaluated along the axis of laser polarization ($y = 0$) and (d-f) $F_y$ forces evaluated perpendicular to the axis of polarization ($x = 0$) under peaks of the (a,d) Dipolar, (b,e) Quadrupolar, and (c,f) Octupolar resonance modes.}
\label{5EDQO_Lines}
\end{figure}

While the force field and the stiffness matrix quantify the binding behavior accurately, they are difficult to read. And since the binding happens along the x and y directions, we can plot the line plots of the sections of force from the quiver plot along these directions. We have shown the the force profiles for the 500~nm gold nanoparticle dimer under weak focusing ($w_0 = 3~\mu\text{m}$) across two the axis parallel to the laser polarization vector ($F_x$ evaluated along $y = 0$, shown in Figs. \ref{5EDQO_Lines}a--c) and the axis perpendicular to it ($F_y$ evaluated along $x = 0$, shown in Figs. \ref{5EDQO_Lines}d--f). In this representation, stable trapping configurations are explicitly identified by zero-force nodes ($F_i = 0$) where the local spatial gradient is strictly negative ($\partial F_i / \partial i < 0$), indicating a physical restoring force that pulls the displaced particle back toward equilibrium.

We also calculated the full two-dimensional vector force quiver plots and corresponding directional trap stiffness maps ($dF_x/dx$ and $dF_y/dy$) for the 400 nm AuNP pair. The qualitative behavior is the same as that of 500 nm, and we have shown those in Section S3 of the Supporting Information.

\subsection{Optical binding force landscapes at the interference of two modes}
While the pure resonance peaks establish highly rigid, directional trapping channels dictated by individual dominant modes, the evolution of the optomechanical landscape in the transition spectral windows between consecutive modes introduces a highly intricate physical regime. At crossover wavelengths where the scattering cross-sections of adjacent multipoles intersect with equal weight (such as the dipole-quadrupole or quadrupole-octupole junctions discussed in Fig. \ref{SCS_OBD}b), the total electrodynamic response is governed by the coherent phase interference between the co-excited orders. This electromagnetic superposition induces sharp asymmetries and phase shifts in the scattered near-field, creating vector force profiles that differ fundamentally from the symmetric, highly stable landscapes mapped at the standalone resonance peaks. To evaluate these transition states, we analyze the electrodynamic force fields and directional tracking gradients at the precise crossover coordinates where adjacent multipolar scattering strengths match. The corresponding two-dimensional vector force fields and directional stiffness maps are reported in the Supporting Information section S4.

\subsection{Effect of beam waist on optical binding}
\begin{figure}[h!]
\centering\includegraphics[width=\textwidth]{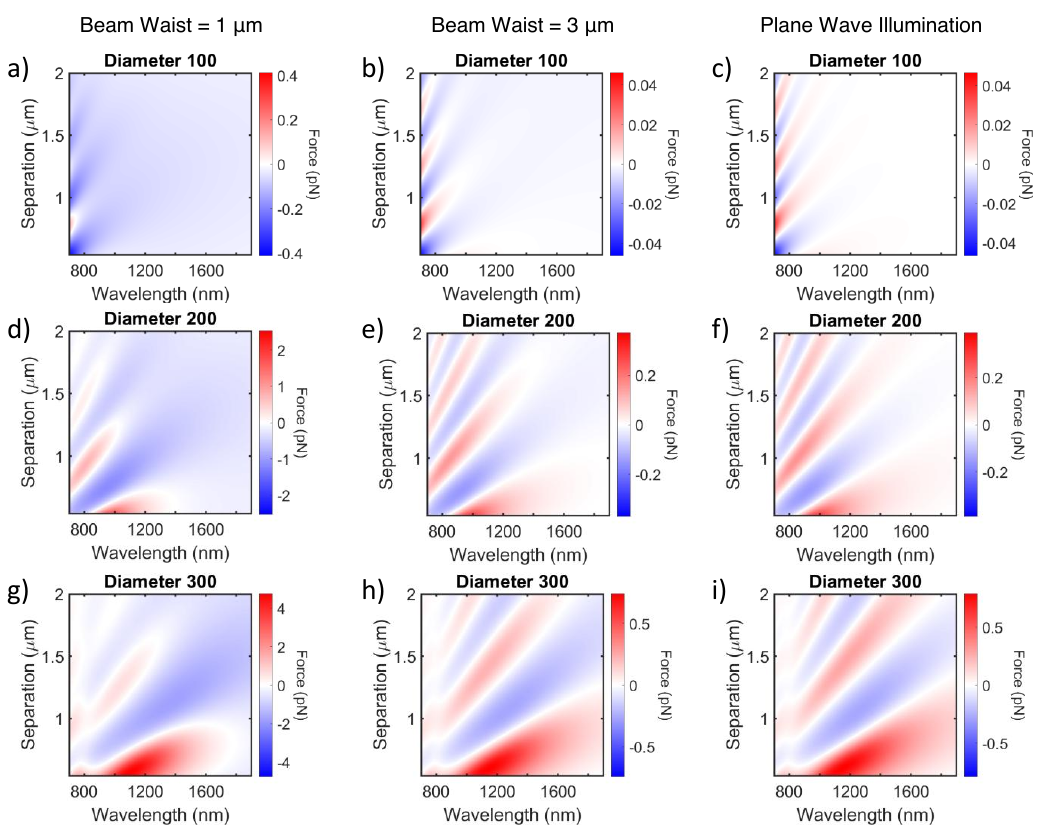}
\caption{Effect of beam focusing on optical binding force landscape. a), b), and c) show the OB force landscape for two 100 nm particles with Gaussian lasers focused to 1 \(\mu m\), 3 \(\mu m\), and plane wave illumination. The force on a 100 nm particle is mostly an attractive gradient force for a 1 \(\mu m\) beam waist. Similarly, d), e), and f) show the OB force landscape for two 200 nm particles. Finally, g), h), and i) show the OB force landscape for two 300 nm particles. The force is oscillatory as the separation increases, but a nonlinear feature emerges at shorter wavelengths. This is due to the emerging prominence of the quadrupolar mode, which peaks at 655 nm for the 300 nm particles.}
\label{123_BW}
\end{figure}

To ensure that the observed force landscapes are driven by interparticle scattering (optical binding) rather than by the trap's external gradient forces, we investigated the effect of beam focusing. A systematic study of particles from 100 to 500 nm enables the observation of the transition from simple dipolar scattering to complex multipolar regimes, where electric quadrupole and octupole resonances emerge. This range is critical to demonstrate how inter-particle binding forces grow to eventually overcome or hybridize with the external gradient forces of the optical trap. Figure \ref{123_BW} presents the force landscapes for particle pairs of increasing diameter $d$ ($100$, $200$, and $300$ nm) under three illumination conditions: strong focusing ($w_0 = 1~\mu$m), weak focusing ($w_0 = 3~\mu$m), and plane wave illumination.

The strong focusing regime ($w_0 = 1~\mu$m, left column) is dominated by the optical gradient force, but it also introduces a fundamental alteration in mode excitation related to displacement resonance\cite{tangMultipoleEngineeringDisplacement2023}. Unlike plane wave illumination, a tightly focused Gaussian beam possesses steep spatial field gradients ($\nabla E$) and decomposes into a spectrum of angular wavevectors. When the beam waist is comparable to the particle diameter (the condition $d/w_0 \approx 1$), these gradients can actively drive higher-order multipole modes, such as the quadrupole and octupole, that are otherwise weakly excited under plane wave illumination\cite{smirnovaMultipolarNonlinearNanophotonics2016}. In experimental conditions, optically bound pairs of particles are inherently off-axis or equidistant from the beam center, and they experience differential excitation effects in which the local phase and amplitude gradients of the beam selectively enhance specific multipolar resonances.\cite{hanGeneralizedLorenzMie2003} Overall, the landscapes in Figure \ref{123_BW}a and \ref{123_BW}d are overwhelmingly attractive, not only due to the macroscopic gradient force pulling particles to the high-intensity center, but also potentially due to this gradient-enhanced multipolar polarizability, which effectively masks the intrinsic oscillatory nature of the optical binding interaction.

In contrast, under weak focusing ($w_0 = 3~\mu$m, middle column) and plane wave illumination (right column), these gradient contributions are suppressed. The $3~\mu$m beam waist is sufficiently large compared to the particle diameter ($d/w_0 \ll 1$) that the wavefront curvature across the particle volume is negligible, effectively behaving similarly to the plane wave case. The force landscapes exhibit clear periodic oscillations between attractive (blue) and repulsive (red) domains as the separation increases. The striking similarity between the $3~\mu$m cases and the plane-wave cases validates our choice of a $3~\mu$m beam waist for the detailed force-landscape calculations. It provides sufficient confinement to localize the particles while remaining flat enough to allow the binding interaction to dominate the force landscape without inducing artificial gradient-driven modes.

Since we mentioned that the binding behavior is anisotropic, we have also calculated the Optical binding force as a function of laser wavelength and separation, parallel to the laser polarization, and reported it in Section S5 of the Supporting Information.

\begin{figure}[h!]
\centering\includegraphics[width=\textwidth]{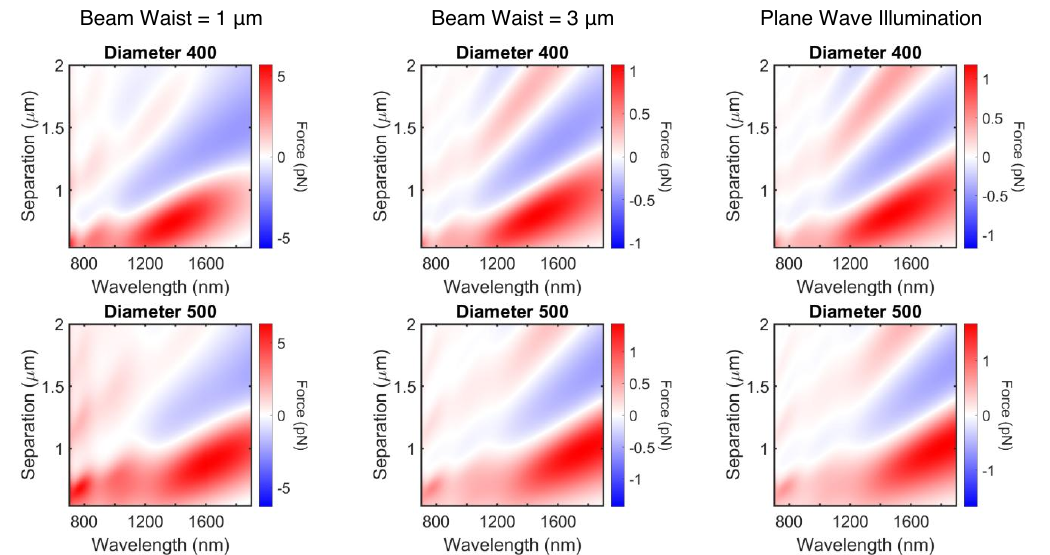}
\caption{Effect of beam focusing on optical binding force landscape. The optical binding force is shown in color as a function of laser wavelength and separation perpendicular to laser polarization. a), b), and c) show the OB force landscape for two 400 nm particles with Gaussian lasers focused to 1 \(\mu m\), 3 \(\mu m\), and plane wave illumination. Similarly, d), e), and f) show the OB force landscape for two 500 nm particles with Gaussian lasers focused to 1 \(\mu m\), 3 \(\mu m\), and plane wave illumination. The force is oscillatory with separation for all cases. The first OB distance, as indicated by the zero force at the lowest separation, increases nonlinearly with wavelength. This is due to the emergence of quadrupolar and octupolar resonances of the particles.}
\label{45_BW}
\end{figure}

Furthermore, these landscapes clearly visualize the onset of intrinsic multipolar effects, as was also shown in Figure \ref{SCS_OBD}c. For the smaller $100$ nm and $200$ nm particles (Figure \ref{123_BW}b-f), the binding points appear as linear, diagonal bands, indicating that the binding distance scales linearly with wavelength ($\lambda$). However, for the $300$ nm particles (Figure \ref{123_BW}h-i), significant distortions appear in the short-wavelength region ($\sim 600 - 700$ nm). These nonlinear features in the force landscape coincide with the emergence of the quadrupolar resonance, which peaks at $655$ nm for a $300$ nm AuNP. This visual confirmation shows that excitation of the quadrupole mode disrupts the simple periodicity of the dipolar binding interaction, fundamentally altering the stability of the array.

%{\color{red} 
As noted in the computational methods, the extreme confinement of the $w_0 = 1\ \mu\text{m}$ case relies on a tightly squeezed Gaussian ansatz. While mathematically exact in its plane-wave propagation, it lacks the full longitudinal polarization complexity of a true high-NA objective lens. Consequently, these specific tight-focusing force landscapes are not intended as quantitatively realistic descriptions of experimental setups, but rather as theoretical limits that demonstrate how macroscopic field gradients can overwhelm intrinsic multipolar binding oscillations.%}

Extending our analysis to the regime of strong Mie resonances, Figure \ref{45_BW} displays the optical binding force landscapes for larger particle pairs with diameters of $d = 400$~nm (a-c) and $d = 500$~nm (d-f). In these size regimes, the scattering cross-sections are significantly larger, leading to inter-particle optical forces that are comparable to or even exceed the gradient forces of the optical trap.

In the strong focusing limit ($w_0 = 1~\mu$m, left column), the force landscape is markedly different from that of the smaller particles. While the $100 - 200$~nm pairs were dominated by the attractive gradient force (appearing almost entirely blue), the $400$ and $500$~nm pairs exhibit strong repulsive regions (red) at short separations, even under tight focusing. This indicates that the scattering-mediated binding forces have grown strong enough to locally overcome the external trapping potential. However, the $1~\mu$m landscape remains a complex convolution of the gradient force and the multipolar optical binding force. Since the particle diameter is now a significant fraction of the beam waist ($d/w_0 \approx 0.4 - 0.5$), the wavefront curvature across the particle is substantial, potentially driving quadrupole and octupole modes with different efficiencies than predicted by standard plane-wave theory.

The intrinsic optical binding behavior is most clearly resolved under weak focusing ($w_0 = 3~\mu$m, middle column) and plane-wave illumination (right column). Here, the force landscape reveals a striking departure from the simple linear scaling observed for smaller dipoles. The zero-force contour, which defines the stable binding separation, does not increase linearly with wavelength but instead exhibits pronounced nonlinear behaviour.

For the $400$~nm pair (Figure \ref{45_BW}b-c), deviations from linearity are visible in the $600 - 800$~nm range. This spectral window corresponds to the interference between the electric quadrupole (EQ, $\approx 796$~nm) and electric octupole (EO, $\approx 619$~nm) modes. The effect is even more dramatic for the $500$~nm pair (Figure \ref{45_BW}e-f). The binding separation exhibits a distinct plateau and a steep rise between 700 nm and 1000 nm. These features align precisely with the spectral positions of the octupolar resonance ($\approx 703$~nm) and the quadrupolar resonance ($\approx 973$~nm).

These nonlinear distortions in the force landscape further support the dependence of optical binding on plasmon modes, as shown in Figure~\ref{SCS_OBD}. As the incident wavelength sweeps across a resonance, the radiation patterns change, fundamentally altering the interference pattern between the incident and scattered fields. Consequently, the equilibrium position is actively pushed and pulled by the dominant multipole mode, resulting in stable binding configurations that are unique to the specific combination of excited modes (e.g., dipole-quadrupole or quadrupole-octupole interference) rather than simple wavelength scaling.

\section{Conclusion}
In this work, we have established a rigorous theoretical framework characterizing the pairwise optical binding landscape of mesoscopic gold nanoparticles beyond the dipolar limit. By systematically exploring the parameter space of particle size ($100$--$500$~nm) and laser wavelength via generalized multiparticle Mie theory, we demonstrated that the spatial topology of stable binding is governed by the specific excitation of higher-order multipole modes.

Our results reveal three critical physical insights into the non-conservative mechanics of Mie-regime optical matter. First, transitioning across individual electric dipole, quadrupole, and octupole resonance peaks does not merely modulate the magnitude of the interaction; it fundamentally reconfigures the trapping topology by rotating the stable zero-force nodes ($\mathbf{F}=0$) and completely transposing the principal axes of the local stiffness tensor. Second, at the crossover transition wavelengths where consecutive multipolar orders possess co-equal scattering cross-sections, coherent multi-modal phase interference triggers severe mechanical trap softening, reducing the directional restoring force gradients ($\partial F_i/\partial i$) to near-zero or destroying stable confinement coordinates entirely. Third, analyzing the illumination focus profiles isolates the precise boundary where scattering-driven fields hybridize with localized beam fields; tight focusing introduces dominating spatial gradient forces and gradient-driven displacement resonances, whereas weak focusing successfully suppresses paraxial gradient contributions, allowing pure scattering-mediated optical binding to dictate the mechanical traps.

By aligning laser wavelengths with specific multipolar resonances, it becomes possible to engineer reconfigurable optical micromachines and programmable metafluids. Furthermore, the nonlinear scaling of binding distance with wavelength suggests new modalities for all-optical sorting of polydisperse nanomaterials based on their multipolar polarizability. Ultimately, this work bridges the gap between fundamental Mie theory and applied nanophotonics, enabling the design of reconfigurable optical matter with tailored mechanical and optical responses.

\begin{backmatter}
\bmsection{Funding}
This work was partially funded by AOARD (grant number FA2386-23-1-4054) and ANRF (grant number ANRF/ARG/2025/002064/PS) to G.V.P.K.

\bmsection{Acknowledgment}
AS acknowledges the Ministry of Education, Government of India, for the Prime Minister's Research Fellowship. We also acknowledge fruitful discussions related to the work with Prof. David Andrews, and our lab members Sumant Pandey, Richard Joseph, and Arindam Maity. %{\color{red}
We finally thank the anonymous reviewers for their thoughtful questions and suggestions that have helped us improve the quality of this manuscript.%}

\bmsection{Disclosures}
The authors declare no conflicts of interest.
 
\bmsection{Data availability} Data underlying the results presented in this paper are not publicly available at this time but may be obtained from the authors upon reasonable request.

\bmsection{Supplemental document}
See Supplementary data appended after bibliography for additional simulation details and results.

\end{backmatter}
%%%%%%%%%% If using BibTeX:
\bibliography{004_MultipolarOB}
\includepdf[pages={1-9}]{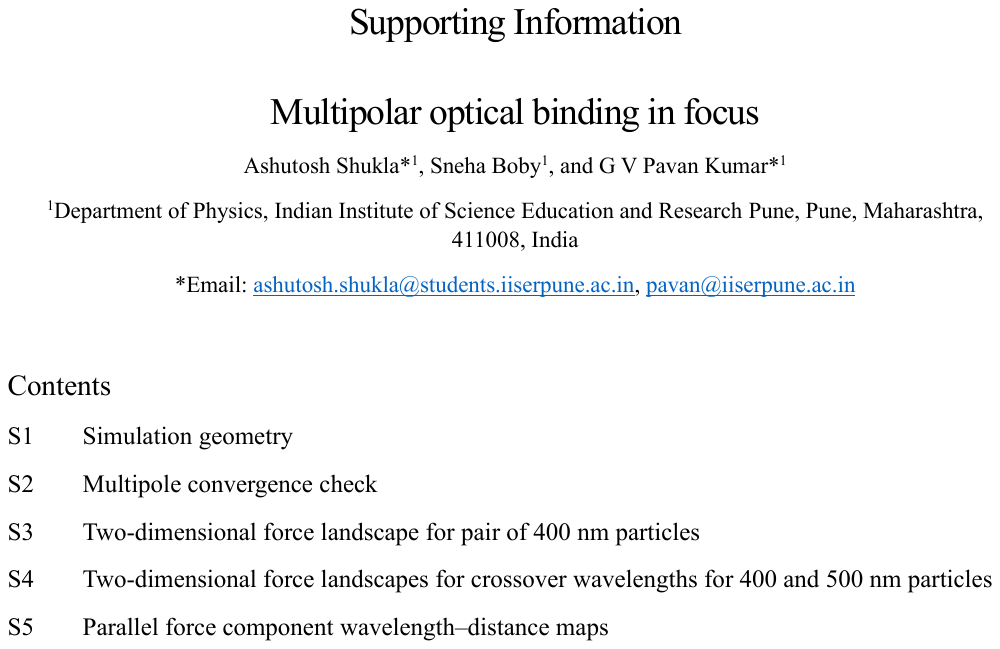}
\end{document}